\documentclass[12pt]{iopart}

 \usepackage{graphicx}
\usepackage{latexsym}
\usepackage{dcolumn}
\usepackage{xcolor}
\usepackage{graphicx, subfigure}
\usepackage{stmaryrd}

\expandafter\let\csname equation*\endcsname\relax
\expandafter\let\csname endequation*\endcsname\relax

\usepackage{amsmath}
\usepackage{graphicx}
\usepackage{dcolumn}
\usepackage{bm}

\begin{document}

\newcommand{\eq}{\begin{equation}}
\newcommand{\eqe}{\end{equation}}
 
\title{Laser induced distortion of band structure in solids: an analytic model}

\author{S. Varr\'o$^{1,2}$, P. F\"oldi$^{2,3}$ and I.F. Barna$^{1,2}$ }
\address{$^1$Wigner Research Centre for Physics of the Hungarian Academy of Sciences \\ Konkoly-Thege Mikl\'os \'ut 29 - 33, H-1121 Budapest,  Hungary\\
$^2$ELI-HU Nonprofit Kft.,  Dugonics T\'er 13, H-6720 Szeged, Hungary \\
$^3$Department of Theoretical Physics, University of Szeged, Tisza L. k\"or\'ut  84 - 86, H-6720 Szeged,  Hungary}
\ead{varro.sandor@wigner.mta.hu}
\vspace{10pt}
\begin{indented}
\item[]December 2018
\end{indented}

\begin{abstract}
 We consider a spatially periodic (cosine) potential as a model for a crystalline solid that interacts with a harmonically oscillating
external electric field. This problem is periodic both in space and time and can be solved analytically using the Kramers-Henneberger co-moving frame. By analyzing the stability of the closely related Mathieu-type differential equation, the electronic band structure can be obtained. We demonstrate that by changing the field intensity, the width of the zero-field band gaps can be drastically modified, including the special case when the external field causes the band gaps to disappear.
\end{abstract}

%

\section{Introduction}
The band-structure plays a fundamental role in our understanding of electronic, transport and also optical properties of crystalline solids. From the theoretical viewpoint, however, the calculation of the band structure for a given material requires computationally expensive numerical approaches. In the following we chose a different route, by focusing on an analytic model that does not aim to reproduce all the properties of a specific crystal, but provides a clear, transparent physical picture for the modification of the band structure in strong, oscillating external fields.

Theoretical methods for obtaining the electronic band structure fall into two categories \cite{yu}: {\it{ab initio}} and empirical approaches. Hartree-Fock, Configuration Interaction (CI) or Density Functional Theory (DFT) methods calculate the electronic structure from first principles, i.e., without the need for any empirical fitting parameters.
Practically, these methods utilize a variational approach to calculate the ground state energy of a many body system, which is defined at the atomic level. Such calculations today are performed using $\sim$ 1000 atoms and, consequently, they are computationally expensive. In contrast to {\it{ab initio}} methods, there are approaches (such as the Orthogonalized Plane Wave (OPW) \cite{her} method, tight-binding \cite{chad} approximation, $k \cdot p$ theory \cite{lutt}, or local \cite{cohen} and non-local \cite{non-local} empirical pseudopotential method (EPM)) that strongly involve empirical parameters to fit experimental data.  (E.g., the band-to-band transitions at specific high-symmetry points that are derived from optical absorption experiments.) The appeal of these methods is that the electronic structure can be easily calculated by solving a single-electron Schrödinger wave equation.
 
Here we concentrate on the fundamental physical mechanisms, and consider a one-dimensional lattice of positive ions as a model. According to Bloch's theorem \cite{bloch}, the solution of the Schr\"odinger equation for a periodic potential can be written as $
   \psi ( x ) = e^{ i k x}  u ( x ),$
where $u(x)$ is a periodic function which satisfies $u(x + \Lambda) = u(x),$ with $\Lambda$ being the lattice constant. As it is known, the band structure corresponding to specific periodic potentials can be obtained, the example that appears in most textbooks is the Kronig-Penney model (see e.g.~\cite{asch} and also \cite{kolm} for more recently obtained results), but the case of the cosine potential can also be treated analytically.

Tzoar and Gersten \cite{tzoar} were the first who examined the band structure of the Dirac comb limit of the Kronig–Penney model in intense laser fields.
Later, Faisal and Genieser \cite{faisal} calculated the exact dispersion relation in a strong external field.
High-order harmonic generation (HHG) is also possible on periodic structures as was shown by Faisal and Kaminsky \cite{fai}. Questions related to the choice of the electromagnetic gauge for the description of the process of HHG were also considered \cite{f17}.
Masovic {\it{et al.} }\cite{mas} calculated the band structure of a solid in external laser field where the one-dimensional Kronig–Penney
model was chosen and the influence of the laser field was taken into account by
a train of periodic Delta-kick pulses. To calculate statistical entropy measure or the Fisher-Shannon information \cite{jamie} of this kind of potentials is also an interesting question. For recent results based on Keldysh cycle-averaged nonperturbative approach, see \cite{GS18}. 

Discussion of additional periodic potentials, including the detailed calculations of bandgaps and the level splittings of the eigenvalues for the one-dimensional Schr\"odinger equation, can  be found in \cite{kirs}.

In the following we consider a cosine potential as a model for crystalline solids and discuss the interaction with an external harmonic electric field.
As we show, the stability chart of the corresponding Schr\"odinger equation (as an ordinary differential equation) can be obtained using the co-moving Kramers-Henneberger \cite{kram,hen} frame. The analysis of the stability chart shows how the external field modifies the band structure.

\section{Model and earlier results}

In 1928, Strutt \cite{strutt} solved the Schr\"odinger wave equation for an electron moving in a periodic, cosine potential. He gave the solutions for two and three dimensional lattices as well. Later several authors applied similar models to investigate the band structure of electrons in crystals \cite{bethe,morse,peierls,slater,seitz}. 
Note that this potential has the appealing property that around each minimum, it can be expanded into power series representing a harmonic oscillator for small amplitudes. Thus the wave functions of tightly-bound states will be like the solutions of the oscillator problem and can be written in terms of Hermite functions.
On the other hand, the higher lying energy levels of the problem are similar to plane waves.

In the following we give a brief overview of Strutt's calculation.
Let's consider the following one-dimensional potential:
$
V(x) = U_0\cos(kx),
\label{egyes}
$
where the spatial amplitude $U_0$ and the wave number $k$ are parameters.  
The corresponding single-particle, time independent Schr\"odinger equation reads:
\eq
-\frac{\hbar}{2m} \varphi''(x) + U_0 cos(kx)\varphi(x) = \epsilon \varphi(x),
\label{maty}
\eqe
where the prime denotes $\partial/\partial x.$
Using the notations $a = \frac{8m\epsilon}{\hbar^2k^2}$  and
$q = \frac{4m U_0}{\hbar^2 k^2},$ the solutions to Eq.~(\ref{maty}) can be written in a closed form as
\eq
\varphi(x) = A e^{-\mu \frac{kx}{2}}\Phi\left(\frac{kx}{2}\right) +  B e^{\mu \frac{kx}{2}}\Phi\left(\frac{-kx}{2}\right),
\eqe
where $A,B$ are the integration constants, $\Phi\left(\frac{kx}{2}\right) $ is a periodic function and $\mu,$ the characteristic exponent,
is usually a complex number. It is woth mentioning here, that since $k$ is real, if $ Re(\mu) \ne 0$ then the solution above diverges as $x \rightarrow \pm \infty.$ For bounded solutions, $\mu,$ which is not arbitrary, should be purely imaginary. For given parameters, the values of $\mu$ can be evaluated using the equation
\eq
\cosh (\pi\mu) = 1 -2\Delta(0)\sin^2 \left(\frac{1}{2}\pi\sqrt{a}\right),
\eqe  	
where $\Delta(i\mu)$ is a tridiagonal determinant of infinite order:

\begin{align}
\Delta(i\mu)  &=
\begin{vmatrix}
. & . & . &  & & \\
\gamma_{-2} & 1 & \gamma_{-2} &  & &  \\
 & \gamma_0 & 1 & \gamma_0 & & \\
 &  & \gamma_{2} & 1 &\gamma_{2} & \\
&   &      . &  .  &  .  & \notag
\end{vmatrix}
\label{det}
\end{align}
with $\gamma_{2r} = q/[(2r-i\mu)^2-a]$ bearing the dependence on $a = \frac{8m\epsilon}{\hbar^2k^2}$ and $q = \frac{4m U_0}{\hbar^2 k^2}$.
Using a different notation, the solutions are the so-called Mathieu C and S functions
\eq
\varphi(x) = c_1C\left(a,q,\frac{kx}{2}\right) + c_2 S\left(a,q,\frac{kx}{2}\right).
\label{maty2}
\eqe
Figure 1 shows the stability chart of the Mathieu differential equation of Eq. (\ref{maty}).
In the black regions, $\mu$ is imaginary and $\varphi(x)$ is stable, while in the white regions $\mu$ is real, $\varphi(x)$ is unstable
and $\Phi(x)$ has a period of $2\pi$ according to \cite{mat1}.
Let us note the detailed properties of the Mathieu functions can be found in the monographs \cite{mat1,mat2,mat3,NIST}.

Mathieu functions play an important role in different fields of physics or in practical applications basically in two main categories:
Boundary-values problems arising from the solution of the two-dimensional wave equation in elliptical coordinates or in stability and initial
value problems. Without completeness, let us mention some applications which are close to the presented problem.
Meixner and Sch\"afke \cite{mat2} used Mathieu functions to solve the quantum mechanical problem of rotation of molecules.
Aly applied \cite{aly} the cosine potential for scattering theory. Investigating relativistic oscillators {\cite{jager}} or phase space properties
{\cite{torres}} also lead to results connected to Mathieu functions.
Maciej and Eberly \cite{mac} found new kind of electron states of hydrogen
in a circularly polarized electromagnetic field where the field-dressed eigenvalues
are addressed with the parameters of the Mathieu functions.
Rozenzweig {\it{et al}} \cite{rosen} proposed a dielectric-loaded resonant laser accelerator where the transversal magnetic modes of the
accelerating waveguide are the Mathieu functions.
Takayama  \cite{taka} investigated the beam-accelerator cavity instability known as beam breakup in a steady-state
free-electron laser in the microwave regime and found that the solutions are related to the Mathieu functions.
In the theory of waveguides \cite{strutt2,milton} for elliptic cylinder coordinates, the solutions are also the Mathieu functions.
\begin{figure}[!h]
\begin{center}
\scalebox{0.6}{
\rotatebox{0}{\includegraphics{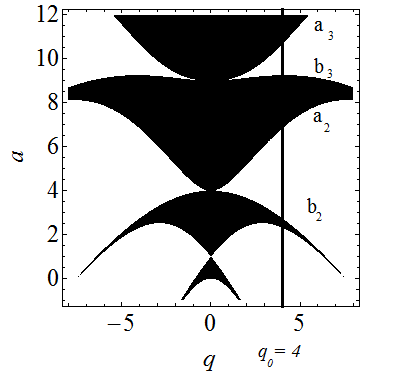}}}
\end{center}
\caption{ a) The stability chart of  the Mathieu equation, where all the parameters are set to unity.  \\
The black regions represents stable and the white regions unstable solutions. According to \cite{mat2,NIST} there are series expansions available
for the limiting stable solutions now called as $b_2  - a_2$ and $b_3 - a_3$.
We will investigate the dynamics of the band gaps  $b_2 - a_2$ and $b_3 - a_3$ in an external field at the point $q_0 = 4$ which is marked by the
black vertical line.}  	
\label{kettes}       
\end{figure}
It is worth mentioning at this point that our experience showed that if the potential Eq. (1) is supplemented with it's first harmonic, $\cos(2kx)$,
the solution is also available in a closed form and can be expressed in terms of the Heun C functions. However, the detailed analysis of this two-color excitation is beyond the scope of the present study.

Power series expansion formulae are available for the limiting curves of the stability regions of Eq. (\ref{maty}) for small and large arguments \cite{mat2, NIST}. In Fig. 1., four such curves, $a_2,a_3,b_2,b_3,$ are present for small arguments  which means $x < 10$. The corresponding polynomials
are the following:

\begin{eqnarray}
\hspace*{-2cm}
a_2(q) = 4 + \frac{5}{12} q^2 - \frac{763}{13824} q^4 + \frac{1002401}{79626240} q^6 + \frac{1669068401}{45 86471 42400} q^8 + ...  \nonumber \\  
\hspace*{-2cm}
b_2(q) = 4 - \frac{1}{12} q^2 + \frac{5}{13824} q^4 - \frac{289}{79626240} q^6 + \frac{21391}{45 86471 42400} q^8 + ... \nonumber \\ 
\hspace*{-2cm}
a_3(q) =  9 + \frac{1}{16} q^2 + \frac{1}{64} q^3 + \frac{13}{20480} q^4 - \frac{5}{16384} q^5 - \frac{1961}{235 92960} q^6 - \frac{609}{1048 57600} q^7 + ... \nonumber \\
\hspace*{-2cm}
b_3(q) =  9 + \frac{1}{16} q^2 - \frac{1}{64} q^3 + \frac{13}{20480} q^4 + \frac{5}{16384} q^5 - \frac{1961}{235 92960} q^6 + \frac{609}{1048 57600} q^7 + ... 
\end{eqnarray}
These are essential results to investigate the dynamics of the band gaps represented by the finite open intervals $(b_2 ... a_2)$ and
$(b_3 ... a_3)$  later on.

\section{The role of the external field}

Let us now consider the external field in dipole approximation, i.e., by neglecting its spatial variance. We assume $ E = E_0 \sin(\omega t),$ where $E_0$ is the
maximal field strength and $\omega$ is the angular frequency of the applied field.
Using the Kramers-Henneberger \cite{kram,hen} co-moving frame the interaction can be reformulated as follows
\eq
V(x)  = U_0\cos(kx +   k\alpha_0 sin[\omega_0 t]).
\eqe
With the dimensionless intensity parameter (or the normalized vector potential)  $\mu = 8.55 \cdot 10^{-10} \sqrt{ I (\frac{W}{cm^2})} \lambda (\mu m),$
the factor of $k \alpha_0 $ can be expressed with the help of the external field wavelength $\lambda$ and the
lattice constant $ \Lambda = 2\pi/k $ in the following way:
$k \alpha_0 = \mu \frac{\lambda}{\Lambda}.$ 

At this point we should mention that for metals in general (like gold or silver) in an 800 nm wavelength Ti:sapphire laser field, the factor $\frac{\lambda}{\Lambda}$  is around 2000, which is the ratio we use in the following calculations.
If $\mu $ is smaller than unity, the non-relativistic description in dipole approximation is valid. (For 800 nm laser wavelength this means a critical intensity of $ I = 2.13 \cdot 10^{18} $W/cm$^2$ for electrons.)

Applying the trigonometric addition formula we get
\eq
\hspace*{-1.5cm}
\cos[kx +   k\alpha_0 sin(\omega_0 t)] = \cos(kx) \cos[k\alpha_0 sin(\omega_0 t)] -
\sin(kx)\sin[k\alpha_0 sin(\omega_0 t)].
\eqe
Using the well-known Jacobi-Anger formulae
\begin{eqnarray}
\cos[k\alpha_0 sin(\omega_0 t)] = \Re[e^{ik\alpha_0 \sin(\omega t)} ]
= \sum_{n=-\infty}^{\infty} J_n(k\alpha_0)\cos(n\omega t) = \nonumber \\
\hspace*{-2cm}
 J_0(k\alpha_0) + \sum_{n=1}^{\infty}[1+ (-1)^2]J_n(k\alpha_0)\cos(n \omega t)
= J_0(k\alpha_0) +2 \sum_{n=2,4,...}^{\infty}J_n(k\alpha_0)\cos(n \omega t)
\end{eqnarray}
and
\begin{eqnarray}
\sin[k\alpha_0 sin(\omega_0 t)] = \Im[e^{ik\alpha_0 \sin(\omega t)} ]
= \sum_{n=-\infty}^{\infty} J_n(k\alpha_0)\sin(n\omega t) = \nonumber \\
\hspace*{-1cm}
 J_0(k\alpha_0) + \sum_{n=1}^{\infty}[1+ (-1)^2]J_n(k\alpha_0)\sin(n \omega t)
= 2 \sum_{n=1,3,...}^{\infty}J_n(k\alpha_0)\sin(n \omega t),
\end{eqnarray}	
the effective laser-dressed potential is obtained by averaging over an optical period
 \eq
\tilde{V}(x) = U_0 J_0(k \alpha_0) cos(kx)].
\eqe
Note that this averaging procedure is valid provided all other relevant processes are considerable slower than the oscillation of the laser field. For narrow-bandgap semiconductor crystals, the optical resonance corresponds to the far infrared regime, which means a time scale which is at least a factor of ten longer than the optical period of the applied laser field. Additionally, since we assumed monochromatic excitation, sub-cycle effects play no role, similarly to \cite{GS18}. 

The solution of the Sch\"odinger equation is now modified as
\eq
 \tilde{\varphi}(x)= c_1C\left[a,\tilde{q} ,\frac{kx}{2}\right] + c_2 S\left[a, \tilde{q} ,\frac{kx}{2}\right],
\label{with_field}
\eqe
where the new potential depth parameter is the following $\tilde{q} =q \cdot J_0\left(\mu \frac{\lambda}{\Lambda} \right) $.
Note that the solutions remain the same, however the stability regions are scaled by the zeroth order modified Bessel function.
For the sake of transparency, we define  $\tilde{\mu}(\sqrt{I}) = \frac{\lambda}{\Lambda} \mu(\sqrt{I}) = C\times \sqrt{I}  $ as the scaled dimensionless
intensity parameter as a natural unit.
For $\lambda = 800 $ nm and $\frac{\lambda}{\Lambda} = 2000,$  the scaling constant has the  value of  $C = 1.368\times 10^{-6}$.
Figure 2 presents the band gaps $(b2 - a2) $ and $(b3 - a3)$ at the original potential depth parameter $q_0 = 4$ (which is
marked by the black vertical line in Fig. 1. ) as the function of the external field intensity.
Note that the first three zeros are at $5.78, 30.48, 74.8,$ in accordance with the behavior of the Bessel functions.
At these intensities the system becomes completely conductive.

This effect is called laser induced insulator-metal transition which has already been experimentally observed in various condensed matter
systems, like doped manganites \cite{hao}, thin films of $C_{60}$ \cite{philips} or in vanadium sesquioxide $V_2O_3$ \cite{vo}.
These are special cases of the more general metal-to-nonmetal transitions which were intensively studied in the last decades \cite{met-nonmet}.
\begin{figure}[!h]
\begin{center}
\scalebox{0.6}{
\rotatebox{0}{\includegraphics{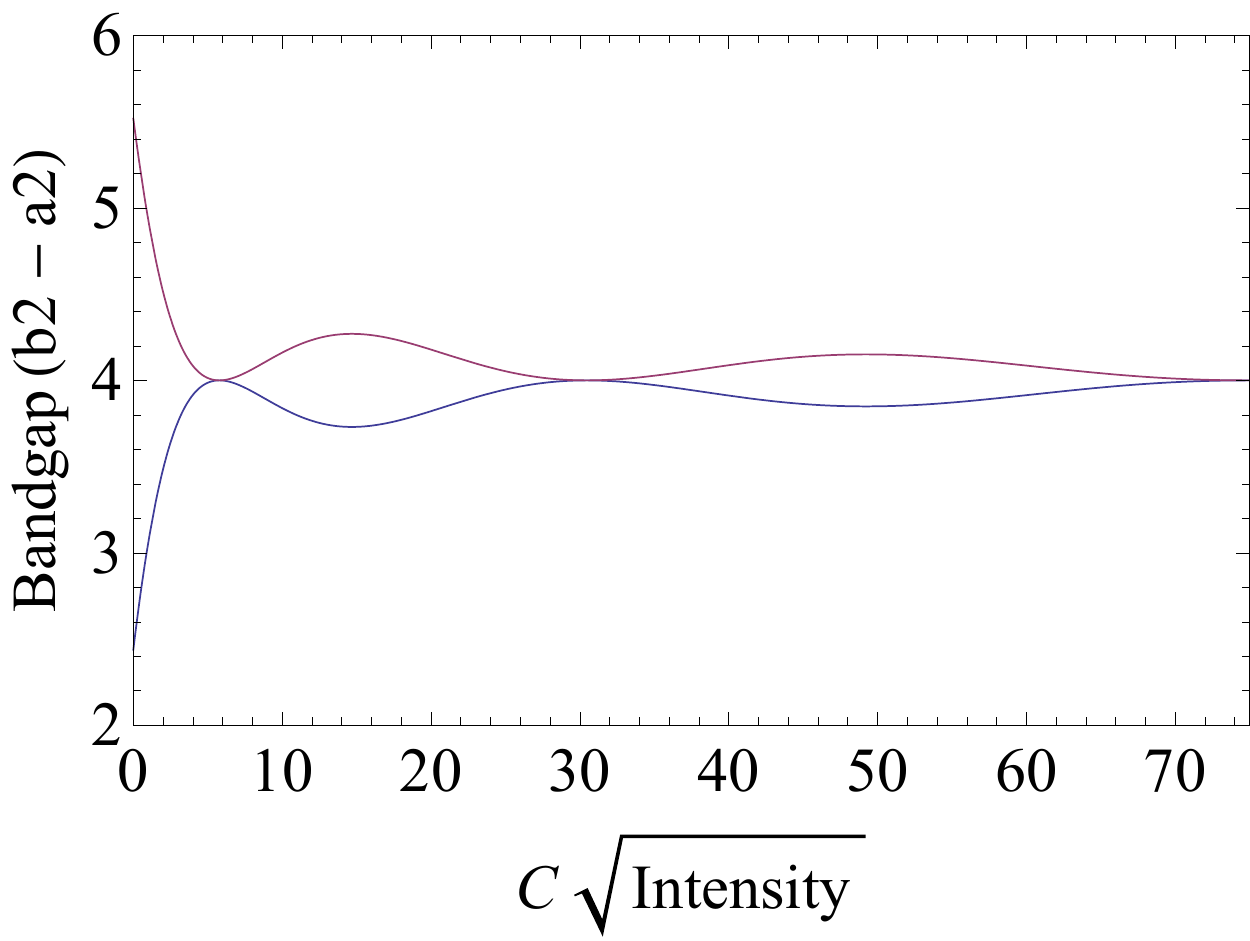}}}
\hspace*{0.3cm}
\scalebox{0.6}{
\rotatebox{0}{\includegraphics{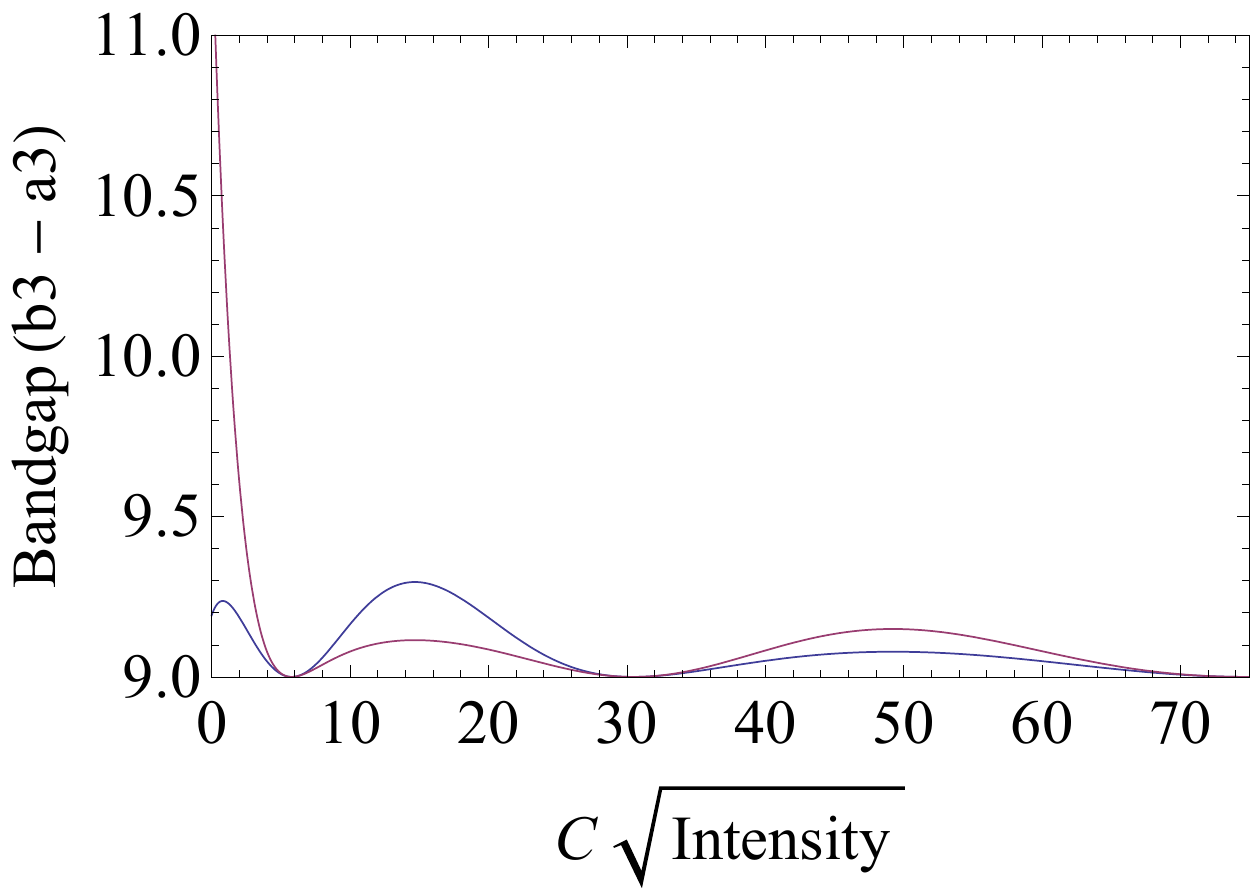}}}
\end{center}
\caption{a)  The width of the $ (b_2 - a_2)$ band gap at the original potential depth parameter $q_0 = 4$
as a function of the field intensity.
The upper line is for $a_2$ and the lower one is for $b_2$.
 b) The same for the next band gap  $ (b_3 - a_3)$.   }  	
\label{kettes}       
\end{figure} 
\section{Summary}
We investigated the dynamics of electron motion in a periodic potential and performed the stability analysis of the corresponding ordinary differential equation.
Working in the Kramers-Henneberger frame, it was shown that in an external harmonic field the stability charts are modified, and
at certain external field intensities the system becomes completely conductive (since all the band gaps disappear).
To emphasize the generality of our approach, let us note that a similar procedure can be followed for the description of the vibrations of diatomic molecules \cite{morse2}.

\section{Acknowledgment}
The ELI-ALPS project (GINOP-2.3.6-15-2015-00001) is supported by the European Union and co-financed by the European
Regional Development Fund. This work was also supported by the European Social Fund under contract EFOP-3.6.2-16-2017-00005. 
\section{References}
 
\end{document}